\documentclass[journal=ancac3,manuscript=article]{achemso}

\usepackage[version=3]{mhchem} 
\usepackage{graphicx}
\usepackage{color}
\usepackage{epstopdf}
\usepackage[colorlinks=true, linkcolor=blue, urlcolor=blue, citecolor=blue]{hyperref}

\newcommand{\CFM}[0]{{
Centro de F\'{\i}sica de Materiales CSIC/UPV-EHU-Materials Physics Center, E-20018 San Sebasti\'an, Spain}}
\newcommand{\DIPC}[0]{{
Donostia International Physics Center, E-20018 Donostia-San Sebasti\'an, Spain}}
\newcommand{\UPVFA}[0]{{
Universidad del Pa\'is Vasco, Dpto. F\'isica Aplicada,  E-20018 San Sebasti\'an, Spain}}
\newcommand{\Ikerbasque}[0]{{
IKERBASQUE, Basque Foundation for Science, 48013, Bilbao, Spain}}

\newcommand{\UAM}[0]{{
Facultad de Ciencias, Universidad Aut\'onoma de Madrid, 28049 Cantoblanco Madrid, Spain}}
\newcommand{\INC}[0]{{Instituto  "Nicol\'{a}s Cabrera", Universidad Aut\'{o}noma de Madrid, 28049 Madrid, Spain}}
\newcommand{\IFIMAC}[0]{{Condensed Matter Physics Center (IFIMAC), Universidad Aut\'{o}noma de Madrid, 28049 Madrid, Spain}}

\author{Amjad Al Taleb}
\affiliation{\UAM}
\author{Frederik Schiller}
\affiliation{\CFM}
\author{Denis V. Vyalikh}
\affiliation{\DIPC}
\alsoaffiliation{\Ikerbasque}
\author{José María Pérez}
\affiliation{\UAM}
\author{Sabine V. Auras}
\affiliation{\CFM}
\alsoaffiliation{\UPVFA}
\author{Daniel Far\'ias}
\affiliation{\UAM}
\alsoaffiliation{\INC}
\alsoaffiliation{\IFIMAC}
\author{J. Enrique Ortega}
\affiliation{\UPVFA}
\alsoaffiliation{\CFM}
\alsoaffiliation{\DIPC}
\email{enrique.ortega@ehu.es}

\title{Simulating high-pressure surface reactions with molecular beams}
\begin{document}

\begin{abstract}

Using a reactive molecular beam with high kinetic energy ($E_{kin}$) it is possible to speed gas-surface reactions involving high activation barriers ($E_{act}$), which would require elevated pressures ($P_0$) if a random gas with a Maxwell-Boltzmann distribution is used. By simply computing the number of molecules that overcome the activation barrier in a random gas at $P_0$ and in a molecular beam at $E_{kin}$=$E_{act}$, we establish an $E_{kin}$-$P_0$ equivalence curve, through which we postulate that molecular beams are ideal tools to investigate gas-surface reactions that involve high activation energies. In particular, we foresee the use of molecular beams to simulate gas surface reactions within the industrial-range ($>$ 10 bar) using surface-sensitive Ultra-High Vacuum (UHV) techniques, such as X-ray photoemission spectroscopy (XPS). To test this idea, we revisit the oxidation of the Cu(111) surface combining O$_2$ molecular beams and XPS experiments. By tuning the kinetic energy of the O$_2$ beam in the range 0.24-1 eV we achieve the same sequence of surface oxides obtained in Ambient Pressure Photoemission (AP-XPS) experiments, in which the Cu(111) surface was exposed to a random O$_2$ gas up to 1 mbar. We observe the same surface oxidation kinetics as in the random gas, but with a much lower dose, close to the expected value derived from the equivalence curve. 

\end{abstract}

\maketitle
\subsection{Introduction}
\label{Intro}
Heterogeneous catalysis is of enormous practical relevance, since it is used to make chemicals and fuels, as well as to reduce car pollution and waste in industry. About 80\% of the chemical manufacturing processes makes use of heterogeneous catalysts, in one form or another. Successful heterogeneous catalytic reactions are generally governed by three elementary steps that take place at the catalyst surface: chemisorption, surface reactions, and desorption. The first step is often the bottleneck for the reaction to occur, due to the presence of a high barrier for the chemisorption of a relevant molecular species. This limits reaction rates if low or moderate temperatures are required ($<$600 K), forcing industrial processes to be carried out at very high gas pressures, in the 100 bar regime \cite{Ra2020,Chorkendorff2003}. 

The high gas-pressure needed in industrially viable high-activation energy processes is also the main challenge to carry out fundamental research in realistic operando conditions, because it lies orders of magnitude above the values at which surface sensitive analytical techniques perform best. This limitation, which is known as the “pressure gap”, was already discussed by Ceyer \textit{et al.} \cite{Ceyer1987}, who tested the methane dissociation combining molecular beams and electron energy loss spectroscopy, demonstrating, in a qualitative way, that the high pressure requirement could be bypassed by raising the energy of the methane molecule. Such idea has been poorly exploited so far in surface chemistry and catalysis, and in particular, in combination with X-ray Photoelectron Spectroscopy (XPS), which stands out as the most accurate to determine the chemical composition of the surface, and the bonding state of each atom. Standard XPS requires Ultra High Vacuum (UHV, $< 10^{-6}$ mbar), because photoelectrons are intensively attenuated in dense gases, although it can also be performed at Near Ambient Pressure (NAP-XPS, or simply, AP-XPS) by selectively pumping the path travelled by the electron from the surface and inside the spectrometer \cite{Ogletree2002}. Pressures up to $\sim 10$ mbar are standard in AP-XPS setups \cite{Schnadt2020}, and a 1 bar pressure has been reached under extraordinary conditions of high-flux, high-energy and grazing incidence X-rays \cite{Amann2019}. Larger pressures appear unattainable nowadays, hence for XPS a 100-bar pressure-gap still remains. A radically new approach is thus needed to bridge the pressure gap for XPS research. Here we quantitatively evaluate the option of performing XPS in UHV in the presence of a reactive monochromatic molecular beam.

\subsection{Random gases versus molecular beams in XPS experiments}

\begin{figure*}
\includegraphics[width=\textwidth]{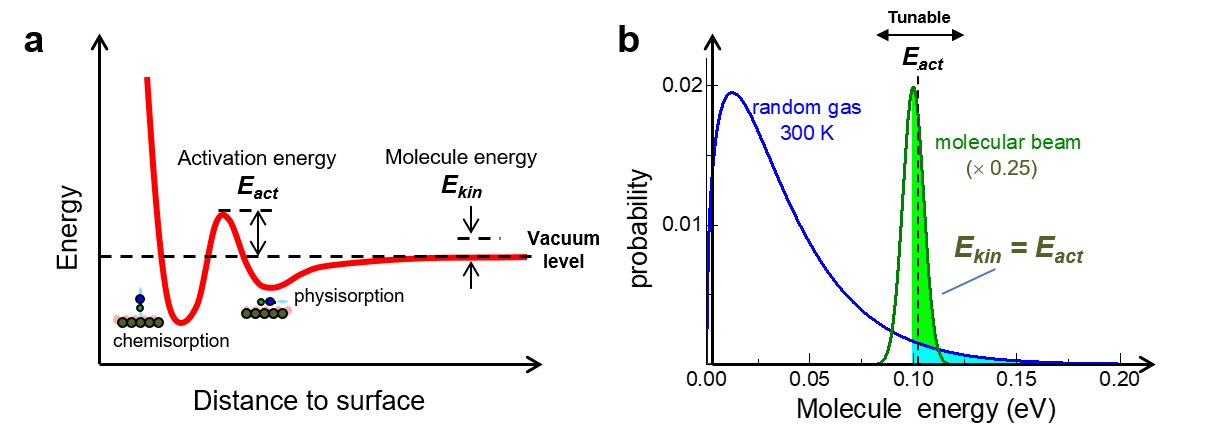}
\caption{\textbf{(a)} Molecule-surface interaction potential for an activated chemisorption process with $E_{act}$ barrier heigth and $E_{kin}$ molecular energy \cite{Juurlink2009}. \textbf{(b)} Normalized (area under the curve = 1) probability distribution of molecular energies fr a random, Maxwell-Boltzmann gas at 300 K, versus a monochromatic, supersonic molecular beam. Assuming $E_{act}$=0.1 eV, and tuning the beam to $E_{kin}=E_{act}$, the blue and green shadowed areas mark the total probability for a molecule to dissociate in the Maxwell-Boltzmann gas and in the molecular beam, respectively.}
\label{fig1}
\end{figure*}

The gas-surface interaction potential for an activated process with $E_{act}$ barrier height and $E_{kin}$ molecular kinetic energy is sketched in Fig. \textbf{1a}. The potential can be more complex \cite{Zaera2017}, but Fig. \textbf{1a} captures the essence of the molecule-surface interaction for chemisorption processes, including those that lead to an immediate dissociation. For instance, O$_2$ adsorbs dissociatively on many metal surfaces at 300 K, typically with low to moderate $E_{act}$, e.g., 0.1-0.2 eV for different Cu facets \cite{Montemore2018}, in contrast to the more inert greenhouse gases, which exhibit larger barriers ($>$ 0.6 eV), as CH$_4$ on Ni and CO$_2$ on Cu surfaces \cite{Juurlink2009,Lustemberg2016a,Yang2020}. For a random gas hitting a surface at 300 K, the molecular kinetic energy is described by the Maxwell-Boltzmann distribution function shown in Fig. \textbf{1b}. Note that the majority of the molecules have relatively low energy. In Fig. \textbf{1b} we assume a hypothetical $E_{act}$= 0.1 eV activation energy, shading in blue the area of the curve with higher $E_{act}$ values. The total probability above $E_{act}$ is given by: 

\begin{equation}
\tau_{gas} (E_{act}) = \int_{E_{act}}^{\infty} \frac{2\pi}{(\pi kT)^{3/2}} \cdot \sqrt{E} \cdot e^{-\frac{E}{kT}} \cdot dE
\end{equation}

Even for the reduced $E_{act}$ = 0.1 eV energy, $\tau_{gas}$ = 0.05, i.e., only 5\% of the molecules will have enough energy to overcome the barrier, chemisorb, dissociate and react \cite{note1}. Moreover, due to the exponential decay of the Maxwell-Boltzmann function, for higher $E_{act}$ the probability would reduce by orders of magnitude.

A molecular beam (MB) is a collimated stream of gas formed by a supersonic expansion from a high source-pressure \cite{Zaera2017}. This process leads to a beam of molecules with well-defined translational energy $E_{kin}$ and narrow energy spread (10\%), as represented by the green Gaussian line in Fig. \textbf{1b}. MB fluxes of $10^{15}\cdot$cm$^{-2}\cdot$s$^{-1}$ are routinely achieved, which corresponds to a $P_{beam}=10^{-6}$ mbar effective pressure on the surface\cite{Kleyn2003}, and energies beyond 1 eV can be obtained, with ample tunability margins. In Fig. \textbf{1b}, both the MB and the random-gas distributions are normalized, i.e., they represent gases of the same molecular density that exert the same pressure on the surface. Assuming a process with $E_{act}$ = 0.1 eV, and tuning the MB energy to $E_{kin}$ = $E_{act}$, half of the molecules (green area) are active, i.e., $\tau_{beam}$=0.5, and the number of active molecules in the MB is 10 times bigger than in the random gas (blue area). On the other hand, molecules with a kinetic energy much larger than $E_{act}$ may not react either, since the excess molecular energy affects its residence time, and may cause its early desorption.

\begin{figure*}
\includegraphics[width=9cm,keepaspectratio]{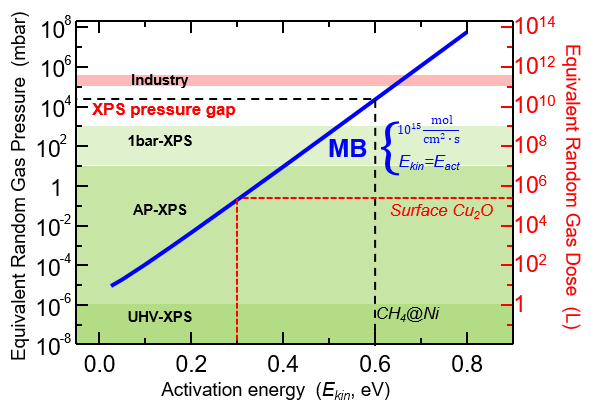}
\caption{Molecular beam energy versus random gas pressure equivalence for gas-surface reactions when tuning the beam to the activation energy $E_{kin}$=$E_{act}$, as deduced from Eq. 1. Such  tuning ability allows the use of standard beams (flux of 10$^{15}$ mol $\cdot$cm$^{-2} \cdot$seg$^{-2}$) to achieve high rates in $E_{act}>$0.6 eV reactions (e.g., CH$_4$/Ni) that require random gas pressures above the XPS regime. The right side scale expresses the random gas dose that is equivalent to the standard molecular beam operating during 1 second (1 L). It is used here to estimate $E_{act}$ from the O$_2$ random gas dose needed to form surface Cu$_2$O (see Ref.\citenum{Liu2022} and the text).}
\label{fig2}
\end{figure*}

From Figure \textbf{1b} it is also immediate to deduce that, for high $E_{act}$ reactions, differences in reactivity between a random gas at 300 K and a regular MB set to $E_{kin}$=$E_{act}$ would be dramatically large. Such difference is rationalized in the "equivalence" curve shown in Fig. \textbf{2}. Following Eq. \textbf{1}, here we define $P_{gas}=P_{beam}/2\tau(E_{act})$, i.e., the pressure of a random gas needed to get the same number of active molecules as in the standard MB with $E_{kin}$=$E_{act}$. This leads to an effective "equivalence" between the pressure of a random gas and the kinetic energy of a MB, as represented with the thick blue line in Fig. \textbf{2}. Analogously, as expressed in the right side scale of Fig. \textbf{2}, the equivalence can also be defined in terms of dose, i.e., the dose of a 300 K random gas needed to provide the same number of active molecules in a standard molecular beam operating during one second, which is equivalent to 1 Langmuir (L, 10$^{-6}$ mbar/s). In the same figure we also depict the pressure ranges at which the different random-gas XPS approaches work. The pressure gap extends from the 1-2 bar limit achieved in Ref. \citenum{Amann2019}, up to the 100 bar range needed, e.g., for the efficient CH$_4$ and CO$_2$ conversion into fuels \cite{Ra2020}. As marked in Fig. \textbf{2} with the dotted line, using a regular MB tuned to $E_{kin}$ = 0.6 eV, within a standard UHV XPS setup, we could simulate operando reactions with random CH$_4$ and CO$_2$ gases at 20 bar, i.e., within the pressure gap, and close to industry values.

The number of high energy, active molecules in a random gas is not the single parameter that governs the reaction probability at a gas-surface interface. The reaction process is dynamic, since the surface does not remain pristine, and hence the activation barrier and the number and nature of active sites changes over the time. Moreover, the chemical nature of the surface may also change because low kinetic energy, inactive molecules may become trapped in physisorbed or chemisorbed states (see Fig. \textbf{1a}). During the CO oxidation on Cu \cite{Eren2016}, at high gas pressures metastable CO condensates form, which disrupt the metal substrate creating active CO/metal-atom clusters that lower the activation barrier. Using the standard $P_{beam}$-pressure MB such high-pressure molecule/metal condensates are not likely to appear. Active surface aggregates are predicted for a variety of molecules and surfaces at high pressures \cite{Xu2023}, but their presence under industrial reaction conditions remains to be proved. In any case, high pressures will always limit the use of standard surface science techniques, such as XPS, to study in detail surface reactions with high activation energies, which can only be done using MBs.

Historically, research on surface reactions exhibiting high activation barriers have greatly benefited from the use of MB sources, which render high reaction rates at $E_{kin} \sim E_{act}$ \cite{Zaera2017,Juurlink2009,Rettner1986,Vattuone1994,Palomino2017,Barratt2019}. In general, MB experiments have been aimed at elucidating reaction kinetics and dynamics, by tuning kinetic energy, vibrational state and incidence angle in the incoming beam of reactants, and probing the same properties in the outcoming gas-phase products \cite{Kleyn2003,Auerbach2021}. To examine the chemical composition of the surface under the simultaneous action of the beam, only infrared spectroscopy has been used \cite{Libuda2002,Libuda2005,Gutierrez2020}, although this technique only probes the chemisorbed species, and not the catalyst surface. XPS could probe both adsorbed molecules and surface atoms, but it has only been used in post-mortem analysis of the surface, e.g., to monitor the oxygen uptake kinetics of Cu surfaces after exposure to O$_2$ beams of variable energy \cite{Moritani2004,Moritani2008,Hayashida2021}. Here, we revisit the Cu(111) surface oxidation process with O$_2$ beams of variable $E_{kin}$ with the aim of testing the random-gas-pressure/molecular-beam-energy equivalence curve shown in Fig. \textbf{2}. Going beyond the total oxygen uptake analysis of Moritani \textit{et al.} \cite{Moritani2004,Okada2007,Moritani2008}, we investigate the separate evolution of the different oxide species, and compare the spectra with recent AP-XPS experiments of random O$_2$ exposure at 300 K and up to 1 mbar. We indeed find the same sequence of oxides, and come to the conclusion that the equivalence works well for the surface oxide species. 

\subsection{Experimental details}

The experiments were performed using a high resolution MB time-of-flight (TOF) apparatus described in detail elsewhere \cite{Barredo2010}. $\rm O_2$ beams have been produced by seeding oxygen using He as a carrier gas. A monochromatic seeded MB was formed by expanding the gas from a 20 mm diameter nozzle operated at 10 bar. The kinetic energy was varied by using different oxygen concentrations (1\% and 20\%) in addition to varying the nozzle temperature between 200 K and 600 K. The mean translational energy of the incoming $\rm O_2$ molecules was determined by measuring the TOF of He atoms in the beam after adjusting for the corresponding mass difference \cite{Scoles}. In this way, we were able to produce $\rm O_2$ beams with kinetic energies from 240 meV up to 1 eV. At a nozzle temperature of 300 K, the corresponding energy spread is ca. 30 \%  for the 20\% $\rm O_2$ beam and   10\%  for the seeded beam with 1\% $\rm O_2$  (see Figs. \textbf{S1} and \textbf{S2} in the Supplementary Information, SI\cite{SI}).  

An electron analyzer (VG CLAM-4) and an X-ray source were inserted into the UHV scattering chamber. Both lie on the scattering plane, defined by the surface normal and the incident MB direction, whereby the sample is placed at the focal point of the analyzer. This arrangement allows measuring XPS spectra immediately after exposing the surface to the MB, simply by rotating the sample 45$^{\circ}$ in the scattering plane.  Clean Cu(111) surfaces were prepared in UHV by repeated cycles of ion sputtering and flash annealing. The surface quality and cleanliness were monitored by measuring a high specular reflectivity to He atoms \cite{Farias2013} and by XPS, prior to each exposure to the $\rm O_2$ beam. 

\subsection{Cu(111) oxidation with low-energy molecular beams}

The oxidation of Cu crystal surfaces has been matter of research since long \cite{Delchar1971}, although still a number of uncertainties remain \cite{Gatinoni2015}. In the presence of a random O$_2$ gas at 300K, it is generally admitted that the cuprous oxide (Cu$_2$O) stoichiometry arises first \cite{Gatinoni2015}, but it is not clear under which conditions the surface oxidation continues into bulk oxide growth, or even further to the formation of a Cu$_2$O overlayer. Structurally, the growth of the oxide has been shown to proceed, in most cases, through the formation of oxide nano-islands which grow and eventually coalesce to form a surface coating layer \cite{Gatinoni2015}, with the morphology and orientation of the surface Cu$_2$O layer notably dependent on the Cu crystal facet \cite{Spitzer1982,Spitzer1982b,Lu2012}. In Cu(100) and Cu(110), the early exposure to the random O$_2$ gas leads to a characteristic Low Energy Electron Diffraction (LEED) pattern, with a sizeable change in surface work function, demonstrating the presence of a well-arranged chemisorbed oxygen layer \cite{Spitzer1982b}. In contrast, O$_2$ adsorption on Cu(111) leads to a rapid disappearance of the LEED ordering, together with a minor change in work function, suggesting a  considerable disruption of the Cu surface and the early incorporation of the oxygen atoms in the subsurface \cite{Spitzer1982}. Higher doses stabilize a bulk-like Cu$_2$O film with similar spectral properties for all crystal facets \cite{Liu2022}. 

\begin{figure*}
\includegraphics[width=14 cm,keepaspectratio]{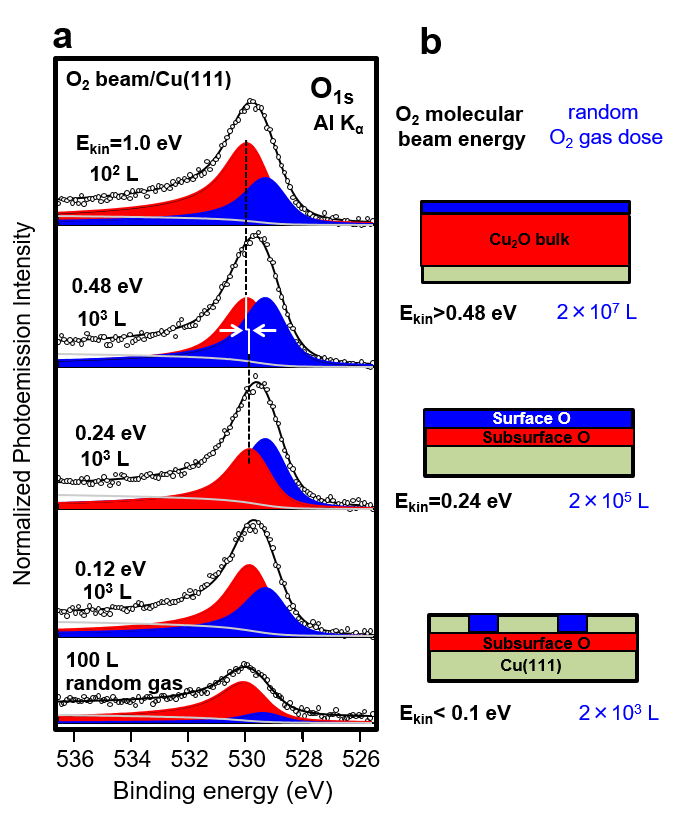}
\caption{\textbf{(a)} O1\textit{s} spectra upon exposure of a clean Cu(111) surface to O$_2$ MB of increasing kinetic energies $E_{kin}$ and low 10$^2$-10$^3$L dose. The changing shape and energy of the feature allows a two-peak fit, which renders the surface O (blue and the subsurface O (red) contributions. The latter grows, shifts (arrows) and widens at $E_{kin}$=0.48-1 eV, indicating the formation of a bulk-like Cu$_2$O phase. The bottom spectrum corresponds to a 30 L random O$_2$ gas exposure. \textbf{(b)} Schematic description of the oxidized Cu(111) surfaces, either after exposing to saturation doses with a molecular beam of variable energy (left scale), or for increasing doses (right scale) of a random O$_2$ gas at 300 K \cite{Liu2022}.}
\label{fig3}
\end{figure*}

Liu \textit{et al.} have recently examined the kinetics of the Cu(111) surface oxidation in AP-XPS experiments within a wide pressure/dose range (UHV-mbar) \cite{Liu2022}. The O1\textit{s} spectrum shows first the buildup of the surface Cu$_2$O oxide with a $\sim$$2 \times 10^5$ L dose, followed by a rapid transition to a bulk-like Cu$_2$O spectrum at higher doses ($2 \times 10^7$ L). Using Liu \textit{et al.} experiment as our random-gas oxidation reference, in Figs. \textbf{3} and \textbf{4} we test the spectral composition of the oxide formed on a Cu(111) surface under the action of a O$_2$ MB with standard flux ($P_{beam}$). In Fig. \textbf{3a} we show the O 1\textit{s} spectra for increasing molecular beam energies and 10$^2$-10$^3$ L doses. Despite the limited resolution (see SI for XPS fitting details and Fig. \textbf{S3}\cite{SI}), a close inspection reveals energy-dependent changes in peak intensity, position and width, which agree with Liu \textit{et al.} data and early O$_2$ beam experiments \cite{Moritani2004,Okada2007,Moritani2008}. Such changes are related to the variable contribution of the different oxide species, shaded with different colors in Fig. \textbf{3a}, and sketched in Fig. \textbf{3b}. More importantly, as indicated in Fig. \textbf{3b}, we observe as a function of the kinetic beam energy the same oxidation sequence obtained when exponentially increasing the dose in the random-gas experiment. Note that in \textbf{3b} we have assumed that the surface oxide is composed of two oxygen species, i.e., subsurface and surface, which are expected to possess different Cu coordination, and hence different O 1\textit{s} emission lines, i.e., at 530 eV and 529.5 eV, respectively. Although Liu \textit{et al.} showed changes in shape and energy in the surface oxide peak similar to those observed in Fig. \textbf{3a}, they did not consider such two-peak contribution. This is otherwise obvious in previous experiments with molecular beams \cite{Okada2007,Moritani2008}, as well as in experiments with random gases when the resolution is improved (see Fig. \textbf{S4} in the SI\cite{SI}). It is clear that the subsurface emission is the first one to emerge with the random gas in Fig. \textbf{3a}, as well as with the smallest dose at any beam energy (see Fig. \textbf{4}) \cite{Okada2007,Moritani2008}. Yet, due to a similar Cu environment, the subsurface oxygen line should lie close to that of bulk Cu$_2$O, i.e., we may expect similar core-level energies. In fact, this is observed in the fit for the subsurface O peak in Fig. \textbf{3a}, which exhibits a $\sim$0.3 eV peak broadening and a small $\sim$0.2 eV higher binding energy shift from $E_{kin}$=0.24 eV to $E_{kin}=$1.0 eV, in line with the literature value for bulk Cu$_2$O, 530.3 eV \cite{Liu2022}.

\begin{figure*}
\includegraphics[width=\textwidth,keepaspectratio]{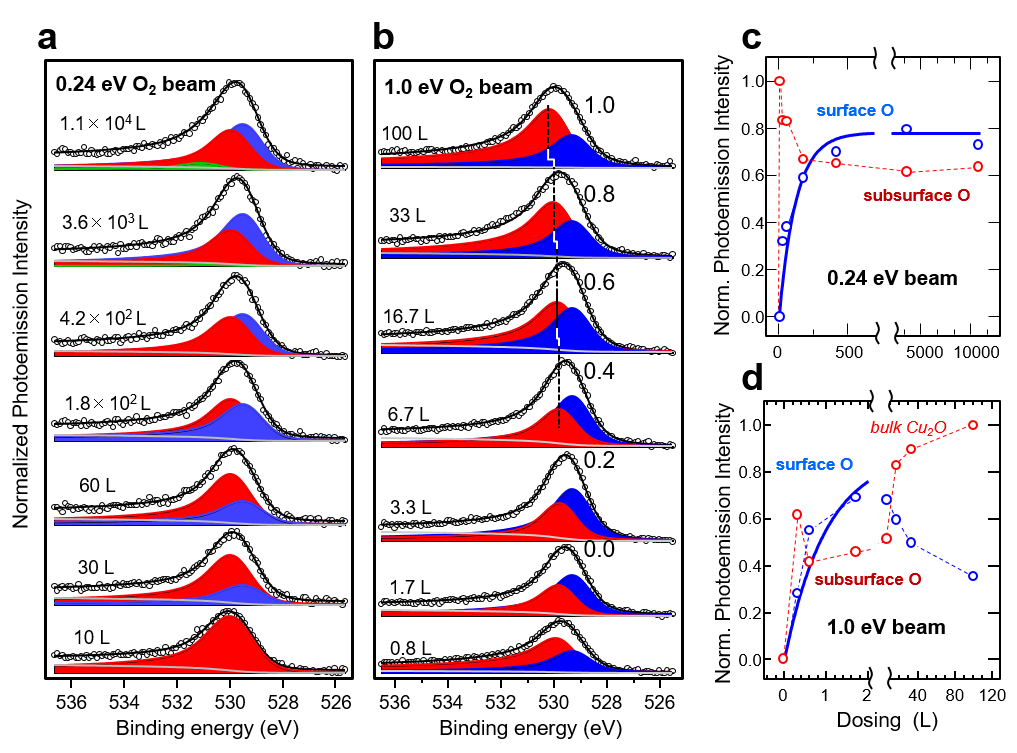}
\caption{Evolution of the O1\textit{s} spectrum for increasing doses of a O$_2$ MB acting on a Cu(111) surface, and tuned to \textbf{(a)} $E_{kin}$=0.24 eV and \textbf{(b)} $E_{kin}$=1.0 eV. Line fitting is performed using two Doniach Sunjic peaks, i.e. surface (blue) and subsurface (red) O, in the same way as in Fig. \textbf{3a}. A third peak (green) is considered in \textbf{a} to account for the small water buildup (hydroxide) at the longest exposures. In \textbf{b}, the subsurface O peak exhibits the high-binding-energy shift that characterizes the emergence of a bulk-like Cu$_2$O layer. The intensity (area under the peak) in each case is represented in panels \textbf{(c)} and \textbf{(d)}. The blue thick lines are fits to surface O data points assuming a simplified model of dissociation probability and two-dimensional growth (see the text).}
\label{fig4}
\end{figure*} 

Spectra in Fig. \textbf{3a} correspond to the doses near saturation at each beam energy, i.e., further exposure to the O$_2$ beam does not result in remarkable spectral changes. We qualitatively conclude, as expressed in Fig. \textbf{3b} sketch, that the surface oxide layer (surface + subsurface O) can be synthesized at $E_{kin}$=0.24 eV using a low beam dose, whereas the bulk-like Cu$_2$O needs beam energies $E_{kin}>$0.48 eV for an efficient growth. Fig. \textbf{4} offers a more quantitative assessment of the Cu(111) oxidation kinetics with beams set to $E_{kin}$=0.24 eV (Fig. \textbf{4a}) and $E_{kin}$=1 eV (Fig. \textbf{4b}). In agreement with the rather small O$_2$ dissociation barrier on clean Cu surfaces \cite{Montemore2018}, we observe the sharp appearance of subsurface O at both energies, prior to the growth of the surface O layer. As the latter saturates, we also notice a slight 30\% drop of the subsurface O peak in both panels, probably due to damping of the photoelectron intensity. The bulk-like Cu$_2$O signal is only observed with $E_{kin}$=1 eV. 

Following the same fitting protocol used in Fig. \textbf{3a}, we obtain the contribution of each species for a given MB dose. The respective "kinetics" at $E_{kin}$=0.24 eV and $E_{kin}$=1 eV are shown in Figs. \textbf{4c} and \textbf{4d}, respectively. As in the random gas \cite{Liu2022}, a three-step process is clearly visible. First, the immediate saturation with subsurface O at the lowest dose, followed by the exponential growth of surface O, and, finally, by the steep rise of the bulk-like Cu$_2$O signal, although the latter only appears, within the limited dosing of the present experiment, with $E_{kin}$=1 eV. The thick lines in Figs. \textbf{4c} fit the surface O data points assuming a simple two-dimensional (2D) growth, i.e., a high dissociation probability upon impact on the portion of the surface that remains "uncovered" with the growing (surface O) species:

\begin{equation}
\dot{\Theta}(t) = K \left[1 - \Theta\right]^n         
\end{equation}
where $K$ stands for the dissociation probability on the "free" surface (saturated with subsurface O), and ${\Theta}$ represents the oxygen coverage, here represented by the surface O peak intensity normalized to its maximum value ${\Theta}^{max}$. In the 2D scenario, $n$=2, since the O$_2$ molecule dissociation requires two contiguous, free adsorption sites, one per each O atom. Then, the integration of Eq. \textbf{2} leads to: 

\begin{equation}
\Theta (t)={\Theta}^{max} \left(\frac{Kt}{1+Kt}\right)       
\end{equation}

The fit to Fig. \textbf{4c} data using Eq.\textbf{3} renders a small $K$=0.023 value, suggesting an effective barrier $E_{act}$ slightly above $E_{kin}$=0.24 eV. In fact, using Liu \textit{et al.}'s surface oxide saturation value of 2$\times 10^{5}$ (Fig. \textbf{3b}), we can estimate $E_{act}$=0.3 eV, as indicated in Fig. \textbf{2}. This value for $E_{act}$ agrees well with the one reported in recent sticking measurements of O$_2$ on Cu(111) using molecular beams \cite{Minniti2012,Zhang2023}. Assuming the probability hypothesis of Fig. \textbf{1b}, in which only those molecules with $E_{kin}>E_{act}$ are active, $K$ would equal the probability ($\tau_{beam}$) about 0.3 eV for a Gaussian distribution with center at $E_{kin}$=0.24 eV and full-width-half-maximum of 0.3$\times E_{kin}$. The algebra is straightforward and leads to $\tau_{beam}$=0.05, i.e., the same order of magnitude of $K=$=0.023. Therefore, despite the crudity of the approaches made, this result supports the central idea here that the number of active molecules above $E_{act}$ governs the reaction kinetics, even in the structurally complex oxidation of the Cu(111) surface. 

For $E_{kin}$=1 eV  (Fig. \textbf{4c}) the surface O intensity curve (blue data) saturates very fast, clearly following a $n<2$ kinetics. This likely reflects the fact that the bulk-like Cu$_2$O signal is also growing in parallel, hence the strict 2D scenario does not hold. Assuming $K=1$ at this energy, the data fit reasonably with $n$=1, meaning that two surface empty sites are no longer needed, since some of the dissociated O atoms incorporate to the bulk directly. Above $\sim$ 20 L, the surface O signal decreases and the Cu$_2$O rises abruptly, in the same way as in the random gas with 2$\times 10^{7}$L dose \cite{Liu2022}. Such quenching of the surface O signal suggests the structural transformation of the Cu$_2$O layer, with a residual amount of O at lower coordination sites of the surface. In any case, using $E_{kin}$=1 eV the MB enables the formation of bulk Cu$_2$O at doses six orders of magnitude lower than in a random gas experiment.\cite{Liu2022}. 

\subsection{Conclusions}

In summary, we have discussed the combined use of molecular beams and XPS to simulate industrially-relevant gas-surface reactions of high activation energy. Based on the Maxwell-Boltzmann distribution function for free gases at 300 K, we have determined the equivalent dose required in the random gas to achieve the same number of active molecules of a standard molecular beam set to a kinetic energy equal to the activation energy of the surface reaction. This equivalence curve reveals fundamental limitations for AP-XPS to investigate many relevant gas-surface reactions with high activation energy, such as those that involve the activation of stable greenhouse gases. As a way of example, we revisited the Cu(111) surface oxidation with O$_2$ beams of variable energy, and focused on the separate kinetics for different oxygen species. We observe the same sequence of oxygen species as in the random gas oxidation, but with a dose which is orders of magnitude lower. In particular, when the kinetic energy of the molecular beam is set close to the activation energy for the surface oxide layer completion, we prove in a quantitative way the accelerated kinetics predicted with the equivalence curve.  

\subsubsection{Acknowledgements}
We want to express our deep gratitude to Prof. Clemens Laubschat (TU Dresden), who willingly donated all XPS instruments used in our experiments. This work has been partially supported by the Spanish Ministerio de Ciencia e Innovaci\'on under project TED2021-130446B-I00. The COST Action CA21101 is also acknowledged. D.F. acknowledges financial support from the Spanish Ministry of Economy and Competitiveness, through the "Mar\'{\i}a de Maeztu" Programme for Units of Excellence in R\&D (CEX2018-000805-M). SVA acknowledges funding from the European Union’s Horizon Europe Programme under the Marie Skłodowska-Curie grant agreement no. 101066965 CURVEO.

\bibliography{references}

\begin{mcitethebibliography}{40}
\bibitem[not()]{note1}
Here we disregard the small contribution from barrier tunneling at E $<$
  E$_{act}$. \relax
\mciteBstWouldAddEndPunctfalse
\mciteSetBstMidEndSepPunct{\mcitedefaultmidpunct}
{}{\mcitedefaultseppunct}\relax
\EndOfBibitem
\end{mcitethebibliography}

\end{document}


\newpage

\begin{figure*}
\includegraphics[width=\textwidth,keepaspectratio]{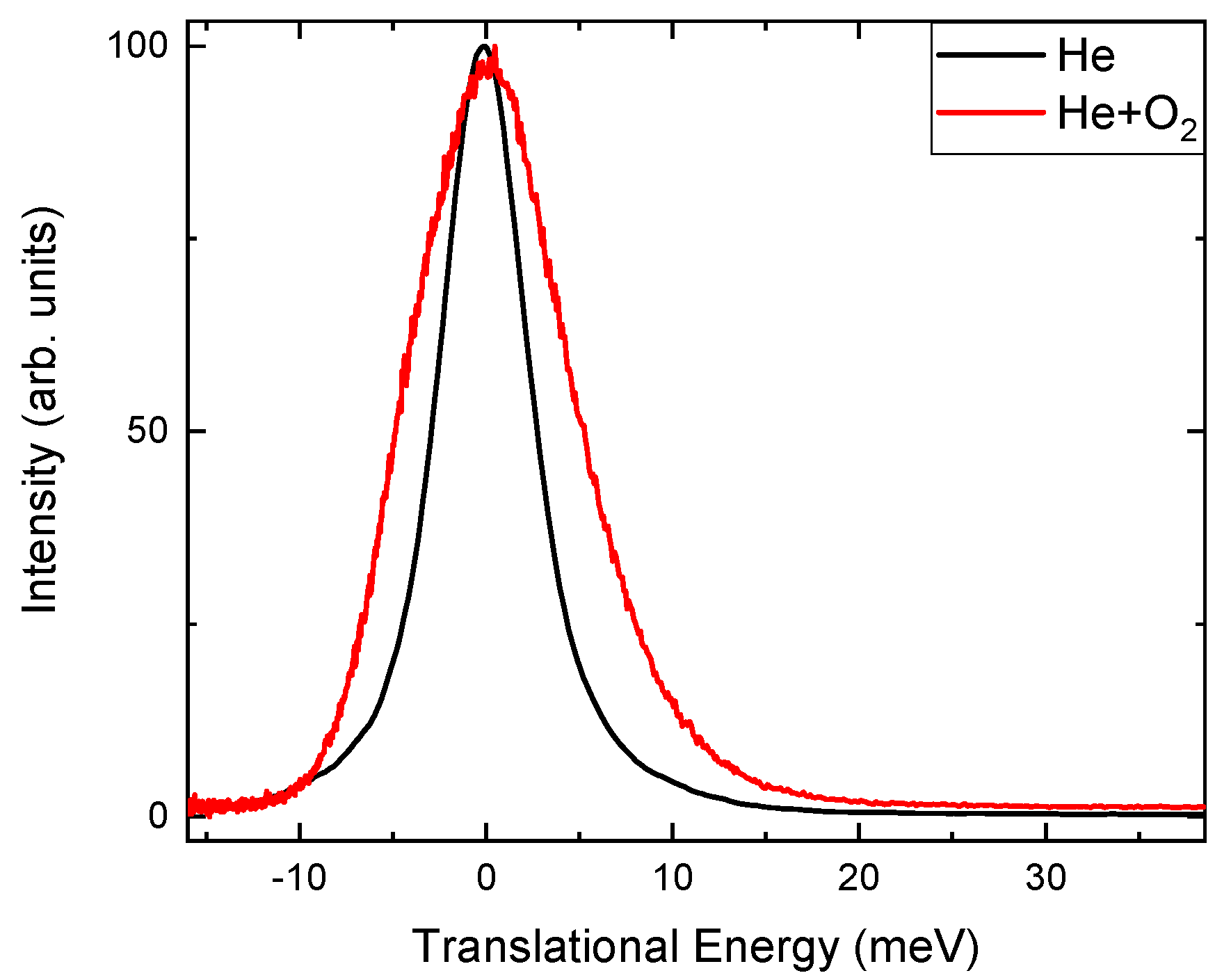}
\caption{\textbf{He/O$_2$ beam:} Kinetic energy distribution of the quasi-elastic peak of a pure He beam (black curve) and a seeded beam with 80\%He-20\%O$_2$ (red curve), measured with a nozzle temperature of 300 K. The corresponding mean energy of He atoms in the beam is 65 meV (black)  and 30 meV (red) and the FWHM 10\% and 33\%, respectively.}
\label{figS1}
\end{figure*}
\clearpage

\begin{figure*}
\includegraphics[width=\textwidth,keepaspectratio]{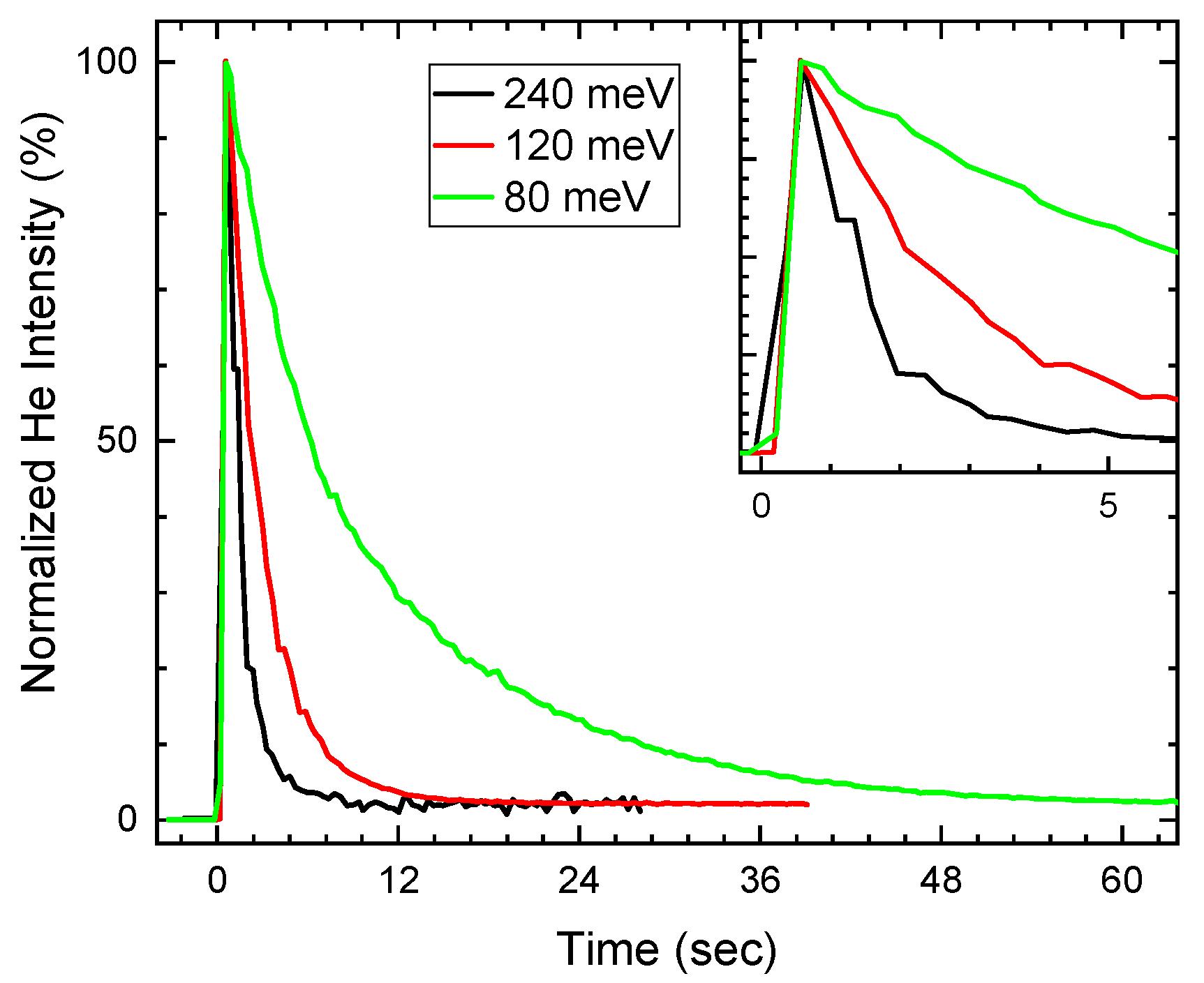}
\caption{\textbf{He/O$_2$ beam:} Attenuation of He specular signal measured from Cu(111) at 300 K using a seeded beam with 80\%He-20\%O2 at an angle of incidence of 45$^{\circ}$ at three incident normal energies: 80 meV, 120 meV and 240 meV. The slope of such a reflectivity curve is proportional to the initial sticking probability of O$_2$ molecules \cite{Poelsema1989}. The increase of ca. a factor of 10 when going from 80 meV (below the dissociation barrier) to 240 meV is consistent with sticking measurements reported by Zhang et al. for the same system\cite{Zhang2023}.}
\label{figS2}
\end{figure*}
\clearpage

\subsection{X-ray photoemission analysis}

X-ray photoemission spectroscopy was carried out by illuminating the sample with Al K$_{\alpha}$ light (h${\nu}$ = 1486.7eV). The source was not monochromatized, therefore the resolution of the light is somehow broad. To estimate the resolution, the measured Fermi edge of the clean crystal at room temperature was fit by a Fermi-Dirac function resulting in a resolution 1.1 eV. For peak fits, Doniach-Sunjic lineshapes \cite{Doniach1970} have been employed together with a Shirley background \cite{Shirley1972} accounting for three peaks. The first peak at lowest energy associates with a Cu bonding configuration at the subsurface (bulk-like), the second one to O atoms at the very Cu surface, and a third peak for a possible hydroxide appearing at higher binding energy. In a first round, all approx. 20 spectra were fit without restrictions. This approach did not give useful results since peak asymmetries, width and background diverted too much. Nevertheless, it gave the predominant peak positions of the subsurface and surface O peaks at 529.6 eV and 530.1 eV, respectively. Furthermore, we observed that hydroxide peaks were only relevant in the long time exposures of the 0.24 eV beam. In a second fit round, the mentioned  peaks were fixed, the width and asymmetry of the subsurface O peak was forced to coincide with the leading surface O peak. From this second round an average width and asymmetry of both peaks were extracted. These two values were then used in a third fitting round to fit all spectra. We noted that the spectra of the higher dosages in the 1 eV beam  were not well fitted, therefore the peak position of the subsurface peak was no longer fixed. This loss of restriction leaded to the peak shift shown in Figs. \textbf{3} and \textbf{4}, characteristic of the transition to a bulk-like Cu$_2$O layer, and re-established the high-quality of the fit.
\pagebreak

\begin{figure*}
\includegraphics[width=\textwidth,keepaspectratio]{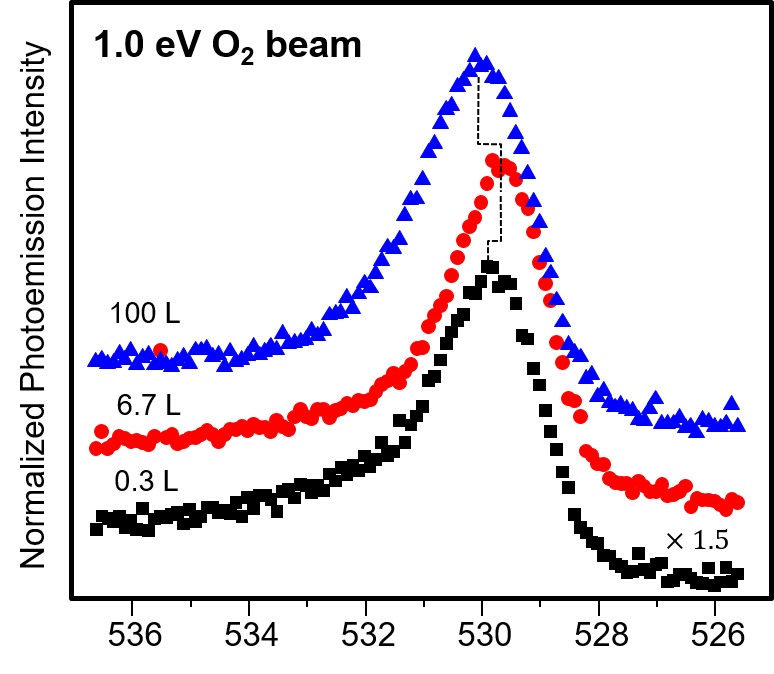}
\caption{\textbf{O 1s spectra at critical Cu oxidation steps:} XPS O1\textit{s} spectra acquired after exposing the Cu(111) to increasing doses of the $E_{kin}$= 1 eV O$_2$ beam. At the appropriate exposure and/or beam energy one can obtain the same successive oxidation steps as in a random O$_2$ gas experiment with exponentially larger doses \cite{Liu2022}. Here we display the representative spectra for the subsurface-O-dominated regime (0.3 L), the complete surface Cu$_2$O layer, and the dominant bulk-like Cu$_2$O phase. Only surface Cu$_2$O has a sizeable presence of the lower-coordinated surface O atom, and hence the whole feature is shifted to the lowest binding energy (see the text). Changes in peak energies and width are in full agreement with earlier O$_2$ beam experiments \cite{Moritani2004,Moritani2008} and recent random-gas oxidation data \cite{Liu2022}.}
\label{figS3}
\end{figure*}
\clearpage

\begin{figure*}
\includegraphics[width=\textwidth,keepaspectratio]{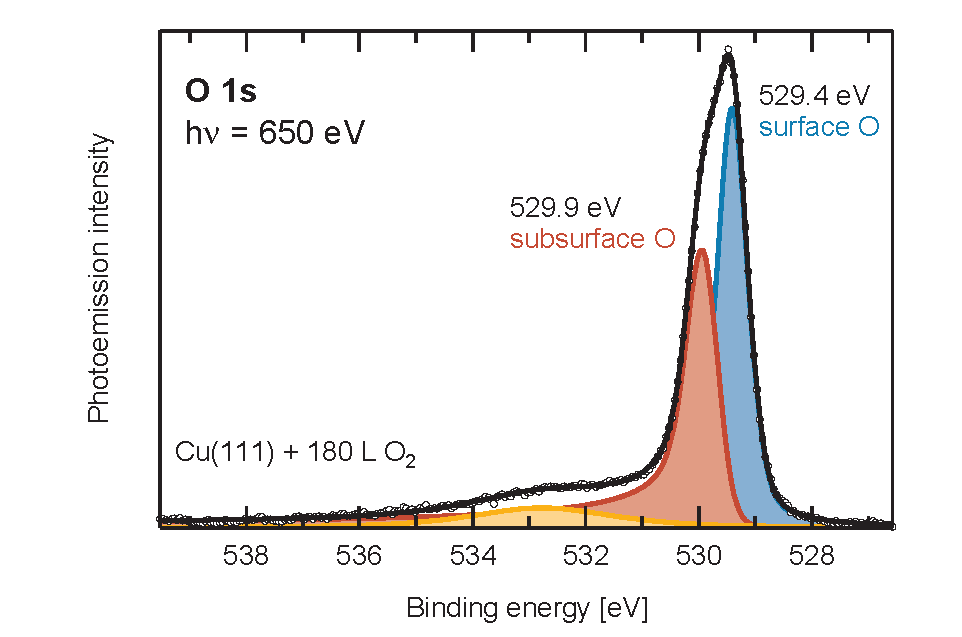}
\caption{\textbf{High resolution surface Cu$_2$O spectrum:} O1\textit{s} spectrum acquired after exposing the Cu(111) surface to 10$^5$ L at 300K. The data have been taken at the IOS beam line of the National Synchrotron Light Source (NSLS-II) in Brookhaven (USA). The excellent energy resolution allows a clear visualization of two different contributions to the spectrum, i.e., those from subsurface and surface O. The shift in the core level energy is readily explained by the high and low coordination of O with Cu atoms at the subsurface and the surface of the Cu$_2$O layer, respectively. Peak energies are in full agreement with the data presented in Fig. \textbf{S3} and the main text.}
\label{figS4}
\end{figure*}

\bibliography{references}

\clearpage